# ENOBIO – FIRST TESTS OF A DRY ELECTROPHYSIOLOGY ELECTRODE USING CARBON NANOTUBES

Giulio Ruffini, Stephen Dunne, Esteve Farrés, Paul C. P. Watts, Ernest Mendoza, S. Ravi P. Silva, Carles Grau, Josep Marco-Pallarés, Lluís Fuentemilla and Bjorn Vandecasteele

*Abstract*—We describe the development and first tests of ENOBIO, a dry electrode sensor concept for biopotential applications. In the proposed electrodes the tip of the electrode is covered with a forest of multi-walled Carbon Nanotubes (CNTs) that can be coated with Ag/AgCl to provide ionic–electronic transduction. The CNT brush-like structure is to penetrate the outer layers of the skin improving electrical contact as well as increase the contact surface area. In this paper we report the results of the first tests of this concept—immersion on saline solution and pig skin signal detection. These indicate performance on a par with state of the art research-oriented wet electrodes.

## I. INTRODUCTION

Fatigue, sleepiness and disturbed sleep are important factors in health and safety and there is considerable interest in developing tools for unobtrusive monitoring of these states. Within the EU FP6 Integrated Project SENSATION [1], sensors, algorithms and systems are being developed for the unobtrusive monitoring of physiological indicators related to sleepiness and stress. In this paper we discuss ENOBIO, a dry electrophysiology sensor employing nanotechnology.

As discussed in an earlier paper [4], electrophysiology electrodes are in high demand in modern clinical and biomedical applications (e.g., electrocardiography, electroencephalography, and electrical impedance tomography)—both research and clinical. The electrode realizes a critical transduction task, and measurement electronics equipment is likely to display misleading artefacts if the electrode concept is defective. Standard electrodes used in high quality low amplitude applications (such as EEG) typically require skin preparation and application of electrolytic gel. This requirement results in longer application times (up to several minutes per electrode) and long stabilization times (diffusion of the electrolytic gel into the skin). In addition, the gel-skin and gel-electrode interface are sources of electrochemical noise.

The ENOBIO sensor is a new type of dry electrode for electrophysiology, to be used to measure EEG or other bio-potential signals. The aim of this electrode design is to eliminate the skin preparation and gel application needed with traditional electrodes, hence simplifying the recording of EEG or other bio-potentials, reducing noise and greatly improving wearability. One of the novel aspects of this electrode is the electrode-skin interface, which consists of a large number of Carbon Nanotubes (CNTs) forming a brush-like structure. This nano-structure will provide a stable low noise electrical interface of low impedance. The CNT structure has been designed to barely penetrate the outer skin layer which consists of dead skin cells and is known as the S*tratum Corneum (SC)* (see Figure 1). The CNT structure will not penetrate so far as to come into contact with nerve cells—resulting in a comfortable and pain free interface. This, and the CNT small diameter, will also hinder possible infection phenomena.

Giulio Ruffini is with Starlab Barcelona S.L., C. de l'Observatori, s/n, Barcelona, 08035 Spain (phone: +34 93 2540366; fax: +34 93 2126445; e-mail: giulio.ruffini@starlab.es).

Stephen Dunne and Esteve Farrés are with Starlab Barcelona S.L., Barcelona, 08035 Spain.

Paul C. P. Watts and Ravi Silva are with the Nanoelectronics Center, Advanced Technology Institute, University of Surrey, GU2 7XH UK.

Ernest Mendoza was with the Nanoelectronics Center, Advanced Technology Institute, University of Surrey, GU2 7XH UK.

Carles Grau, Josep Marco-Pallarés and Lluís Fuentemilla are with the Neurodynamics Laboratory, Department of Psychiatry and Clinical Psychobiology, University of Barcelona, P. de la Vall d'Hebron 171, 08035 Barcelona, Catalonia, Spain.

Bjorn Vandecasteele is with IMEC, Kapeldreef 75, B-3001 Leuven, Belgium.

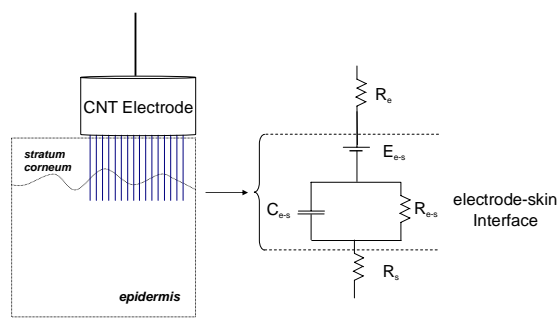

**Fig. 1. Electrode interface equivalent circuit, showing the inherent electrode resistance, the contact related half-cell potential and parallel capacitance-resistance, and epidermis resistance.**

The mechanical interaction of CNTs and skin is poorly understood because the mechanical properties of the skin at nanoscale are difficult to estimate, but the initial sensor paradigm that we will investigate assumes CNTs penetrating or puncturing the outer skin layer to a depth of 10-15 µm.

Another potential paradigm is the case of coupling enhanced by the increase in contact surface area. In this case shorter CNTs shall be used. Both paradigms shall be investigated. Finally, we recall that the electrode must transduce ionic currents to electronic currents; the initial penetrative design targets a nonpolarisable electrode. Therefore, an Ag/AgCl coating will be targeted. The Ag/AgCl coating transduces current via a redox reaction as in traditional "wet" electrodes. A polarizable version without coating will first be tested.

## II. CNT INTERFACE

### A. CNT characteristics

To develop an electrophysiological sensor, multiwalled CNTs were chosen because they meet good conductivity requirements and are stronger than single-walled carbon nanotubes. The architecture of the CNTs arrays was adjusted to achieve penetration of only the outer layer of the skin and to avoid pressure distribution commonly known as the "bed of nails" effect, which could hinder penetration. The catalyst used to grow the CNTs was iron.

### B. CNT Interface Growth

The multi wall carbon nanotubes (MWNTs) arrays have been grown on highly doped silicon substrates (Charntec Electronics <1-0-0> N type 0.8 - 0.15 $\Omega$.cm) using plasma enhanced chemical vapour deposition (PECVD) of acetylene over an iron catalyst. Generally, a *ca.* 10 nm iron film was sputtered onto the silicon wafer immediately after etching the native oxide by immersion in hydrogen fluoride. The growth process consists of heating the substrate to 650 ºC for 20 min in vacuum in order to break the Fe film into small islands between 50 - 100 nm in diameter. During the growth process the substrate was maintained at 650 ºC and acetylene was introduced to the chamber at a concentration of 5.0 % with $H_2$ as the carrier gas. The as-grown MWNTs have diameters *ca.* 50 nm and the length depends on growth time. Typically, the length of the MWNTs for a 15 minute growth process is *ca.* 20 - 30 μm.

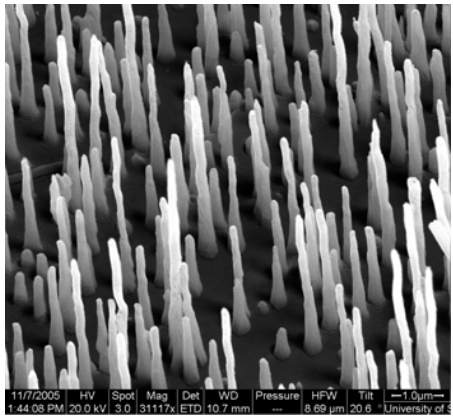

**Fig. 2. CNT array**

An underlying drawback for the use of CNT arrays as sensing platforms is the fact that the nanotubes are relatively weakly bound to the silicon substrate and can become detached; this is an important consideration in light of possible biocompatibility issues associated to nanostructures. The CNT arrays are protected against nanotube detachment by securing them to the substrate by means of a 400 nm thick polymer support structure, while ensuring that the main body and ends of the CNTs remain electrically active.

We have also coated the CNT surfaces with an ion exchange substance, a silver/silver chloride (Ag/AgCl) bilayer, in order to improve transduction in biological fluids. This was accomplished by coating the nanotubes with silver by thermal evaporation, after which chlorine was electrodeposited to form AgCl. The successive steps were characterized by electron energy dispersive X-ray spectroscopy and the bio redox potentials were evaluated by cyclic voltametry. We are currently investigating the biocompatibility of CNTs with biological systems and the performance of the arrays as biopotential sensors.

## III. ELECTRODE TEST PROTOTYPE I

The development approach we have taken is based on two steps. The first CNT arrays were mounted on state of the art commercial active electrodes (Biosemi Active 2 [2, 3]) and connected to commercial-off-the-shelf research electrophysiology recording equipment in order to be compared to existing "wet" sensors. This approach saves considerable development time while in the proof of concept phase and is entirely satisfactory for testing purposes of the CNT-interface concept.

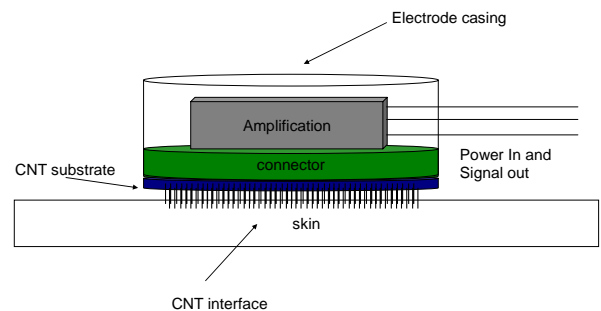

**Fig. 3. ENOBIO Test Prototype I**

Once Prototype I has been validated we shall integrate on-site amplification and produce Prototype II. Prototype II shall be an end to end solution incorporating amplification and wireless data transmission to a base station. Finally the possibility of growing CNTs directly on thinned flexible substrates for improved wearability will be investigated.

## IV. INITIAL RESULTS

The first electrode prototypes used for validation and which we wish to report here have the following characteristics: CNT Diameter: 50 nm; CNT Length: 10-15 μm; Catalyst: Fe; Coating: none.

### A. Results 1 - Noise Comparison

A practical first requirement for a "good" electrode [2] is that it exhibits low noise in the region of 1-2 μV RMS in the ~0.1-100 Hz range when immersed in saline solution. One of our initial tests was a direct comparison of noise spectral density for two ENOBIO Prototype 1 electrodes and two state-of-the-art commercial electrodes in saline solution. The amplitude spectral densities for one of the commercial electrode (from Biosemi) and an ENOBIO electrodes is provided in the following figures (based on 10 takes of 16 seconds each and 0.1 Hz high pass filtering):

As can be seen, the RMS noise measured for the two ENOBIO electrodes is rather similar to that from the commercial electrodes. This is an important check as any electrode with an intrinsic noise level higher than a few μV will perform poorly in EEG applications.

### B. Results 2 – Signal Detection

In order to test signal response in a realistic situation the electrodes were placed on pig skin and a small test signal applied beneath the skin. Pig skin is similar in structure to human skin and provides a good starting point for prototype development.

A commercial electrode was applied to pig skin in the usual manner—using electrolytic gel—while ENOBIO was applied without gel or skin preparation. The following plots compare electrode response to an applied 0.3V/10Hz signal. The results from the comparison are again rather similar.

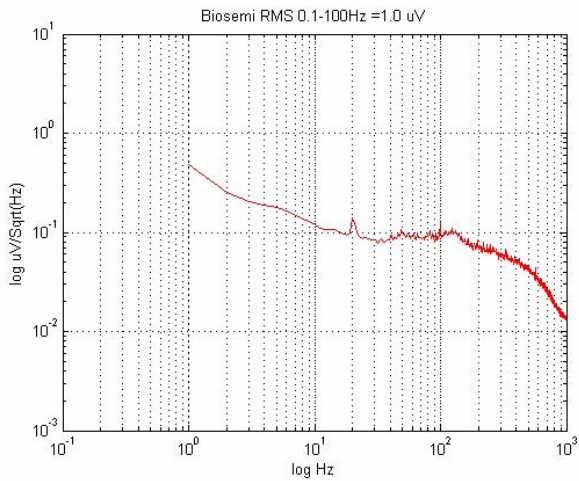

**Fig. 4. Commercial electrode amplitude spectral density with RMS noise of 1.0 μV from 0.1 to 100 Hz**

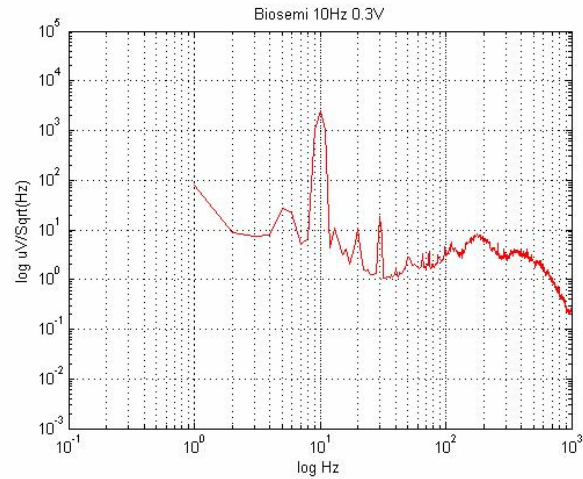

**Fig. 6. Commercial wet electrode 10Hz peak**

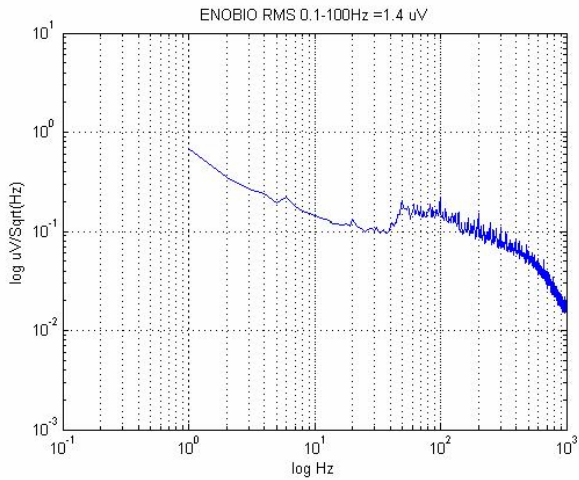

**Fig. 5. ENOBIO electrode amplitude spectral density with RMS noise of 1.4 μV from 0.1 to 100 Hz**

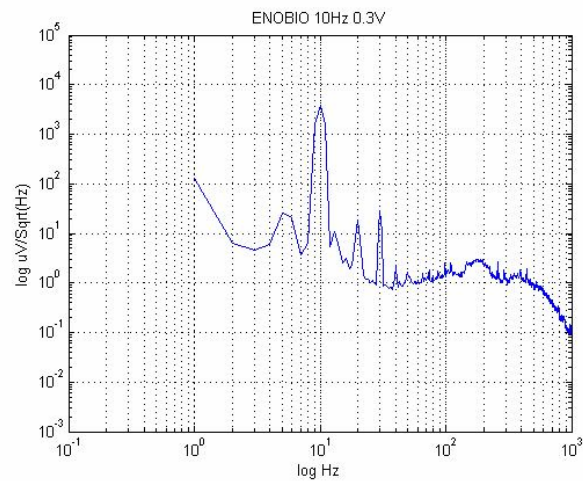

**Fig. 7. ENOBIO dry electrode 10 Hz peak**

This is an important first result as it provides a good indication of the signal detection performance of a dry ENOBIO electrode in comparison to a wet electrode.

## V. ELECTRONIC DESIGN AND FLEXIBLE SUBSTRATE

In parallel to the development of the CNT transduction interface, a new electronic interface or back end has been designed and tested. ENOBIO will be an active electrode with on board amplification. On board amplification is a proven improvement which greatly reduces interference noise [3,4]. In addition, the electrode measurement system will be wireless enabled for use within a Body Area Network (BAN) that is also being developed within the SENSATION project. This infrastructure is based on the Zigbee standard and can handle multi-channel EEG/ECG with sampling rates of up to 500 Hz.

A final step in the process will be the translation of the electronic design to a thinned flexible solution for improved wearable applications. Flexible silicon substrates will be integrated with unencapsulated electronic components in order to improve wearability of the final prototype. These and other developments will be reported elsewhere.

## VI. CONCLUSIONS

The design of a CNT-based electrophysiology electrode is a fascinating and challenging multi-disciplinary exercise involving analysis of requirements, skin and CNT mechanical and electrical properties at nanoscale, electrochemistry and biosafety issues. In a previous publication we analyzed requirements for EEG/ECG/EOG applications and provided the logic for the electrode design, starting from a careful analysis of all the noise sources and measurement requirements [4]. It was concluded that a significant portion of the problems for practical applications of electrophysiology arise from the need for gel application, scrubbing and/or Faraday caging to minimize noise. As discussed, the source of the most difficult-to-handle noise sources is the SC-gel interface. To address this, we have proposed a new concept of dry electrode based on the use of multi-walled carbon nanotubes to penetrate the outer (dead/dry) cell layers of the skin and thus reduce the noise of measurements. This, it was argued, will allow recording of biopotential signals, especially EEG, without the use of gel or skin scrubbing/abrasion. An alternative mechanism for improvement of the electrical coupling is the significant increase of contact surface area provided by CNT arrays.

As we have reported, early results on the design performance are encouraging and indicate that the dry ENOBIO concept will perform favourably in comparison to state of the art research-oriented commercial wet electrodes. The tests carried out so far involve simple immersion on saline solution and signal detection with animal skin. In a future publication we will report results from the first human tests of this concept as well as the overall electronic design.

## VII. ACKNOWLEDGEMENTS

ENOBIO is developed under the FP6 European Project SENSATION (FP6-507231). G. Ruffini is thankful to A. Kasumov and C. Ray for useful discussions.